\documentclass[final]{IEEEtran}
\pdfoutput=1

\usepackage{amsmath, amssymb}
\usepackage{graphicx}
\usepackage{subfigure}
\usepackage[noadjust]{cite}
\newtheorem{theorem}{\textbf{Theorem}}
\newtheorem{remark}{\textbf{Remark}}
\newtheorem{lemma}{\textbf{Lemma}}
\usepackage{color}

\title{
A Framework for Robust Assessment of Power Grid Stability and Resiliency
}
\author{Thanh Long Vu,~\IEEEmembership{Member,~IEEE,} and~Konstantin~Turitsyn,~\IEEEmembership{Member,~IEEE}
\thanks{Thanh Long Vu and Konstantin Turitsyn are with the Department of Mechanical Engineering, Massachusetts Institute of Technology, Cambridge, MA 02139, USA, e-mail: longvu@mit.edu and turitsyn@mit.edu.

}}

\begin{document}

\maketitle
\begin{abstract}
Security assessment of large-scale, strongly nonlinear power grids
containing thousands to millions of interacting components is a
computationally expensive task. Targeting at reducing the
computational cost, this paper introduces a  framework for constructing a robust
assessment toolbox that can provide mathematically rigorous
certificates for the grids' stability in the presence of variations
in power injections, and for the grids' ability to withstand a bunch sources of faults. By this toolbox
we can ``off-line'' screen a wide range of contingencies or power injection profiles, without reassessing
the system stability on a regular basis.
In particular, we formulate and solve two novel robust
stability and resiliency assessment problems of power grids subject to the uncertainty in
equilibrium points and uncertainty in fault-on dynamics.
Furthermore, we bring in the quadratic Lyapunov functions approach
to transient stability assessment, offering real-time construction
of stability/resiliency certificates and real-time stability assessment. The
effectiveness of the proposed techniques is
numerically illustrated on a number of IEEE test cases.
\end{abstract}

\section{Introduction}

\subsection{Motivation}

The  electric power grid, the largest engineered system ever, is
experiencing a transformation to an even more complicated system
with increased number of distributed energy sources and more
active and less predictable load endpoints. Intermittent renewable
generations and volatile loads introduce high uncertainty into
system operation and may compromise the stability and security of
power systems. Also, the uncontrollability of inertia-less
renewable generators makes it more challenging to maintain the
power system stability. As a result, the existing control and
operation practices largely developed several decades ago need to
be reassessed and adopted to more stressed operating conditions
\cite{1705631,turitsyn2011options,5454394}. Among other
challenges, the extremely large size of the grid calls for the
development of a new generation of computationally tractable
stability assessment techniques.

A remarkably challenging task discussed in this work is the
problem of \textit{security assessment} defined as the ability of
the system to withstand most probable disturbances. Most of the
large scale blackouts observed in power systems are triggered by
random short-circuits followed by counter-action of protective
equipments. Disconnection of critical system components during
these events may lead to loss of stability and consequent
propagation of cascading blackout. Modern Independent System
Operators in most countries ensure system security via regular
screening of possible contingencies, and guaranteeing that the
system can withstand all of them  after the intervention of
special protection system \cite{SPS}. The most challenging aspect
of this security assessment procedure is the problem of certifying
\emph{transient stability} of the post-fault dynamics, i.e. the
convergence of the system to a normal operating point after
experiencing disturbances.

The straightforward approach in the literature to address this
problem is based on direct time-domain simulations of the
transient dynamics following the faults
\cite{Huang:2012il,Nagel:2013kf}. However, the large size of power
grid, its multi-scale nature, and the huge number of possible
faults make this task extremely computationally expensive.
Alternatively, the direct energy approaches \cite{Pai:1981dv,
Chiang:2011eo} allow fast screening of the contingencies, while
providing mathematically rigorous certificates of stability. After
decades of research and development, the controlling  UEP method
\cite{Chiang:1994ir} is widely accepted as the most successful
method among other energy function-based  methods and is being
applied in industry \cite{Tong:2010}. Conceptually similar is the
approaches utilizing Lyapunov functions of Lur'e-Postnikov form to
analyze transient stability of power systems
\cite{Hill:1989:LFL:72068.72082, Hiskens:1997Lya}.

In modern power systems, the operating point is constantly moving
in an unpredictable way because of the intermittent renewable
generations, changing loads, external disturbances, and real-time
clearing of electricity markets. Normally, to ensure system
security, the operators have to repeat the security and stability
assessment approximately every 15 minutes.
For a typical power system composed of tens to hundred thousands
of components, there are millions of contingencies that need to be
reassessed on a regular basis. Most of these contingencies
correspond to failures of relatively small and insignificant
components, so the post-fault states is close to the stable
equilibrium point and the post-fault dynamics is transiently
stable. Therefore, most of the computational effort is spent on
the analysis of non-critical scenarios.
This computational burden could be greatly alleviated by a robust
transient stability assessment toolbox, that could certify
stability of power systems in the presence of some uncertainty in
power injections and sources of faults.
This work attempts to lay a theoretical foundation for such a
robust stability assessment framework. While there has been
extensive research literature on transient stability assessment of
power grids, to the best of our knowledge, only few approaches
have analyzed the influences of uncertainty in system parameters
onto system dynamics based on time-domain simulations
\cite{hiskens2006sensitivity, dong2012numerical} and moment
computation \cite{dhople2013analysis}.



\subsection{Novelty}
This paper formulates and solves two novel robust stability
problems of power grids and introduces the relevant problems to controls community.

The first problem involves the transient stability analysis of
power systems when the operating condition of the system variates.
This situation is typical in practice because of the natural
fluctuations in power consumptions and renewable generations. To
deal with this problem, we will introduce a robust transient
stability certificate that can guarantee the stability of
post-fault power systems with respect to a set of unknown
equilibrium points. This setting is unusual from the control
theory point of view, since most of the existing stability
analysis techniques in control theory implicitly assume that the
equilibrium point is known exactly. On the other hand, from
practical perspective, development of such certificates can lead
to serious reductions in computational burden, as the certificates
can be reused even after the changes in operating point.

The second problem concerns the robust resiliency of a
given power system, i.e. the ability of the system to withstand a set of unknown faults and return to stable operating conditions. In vast majority of power systems subject to faults, initial disconnection of power system components is followed by consequent action of reclosing that returns the system back to the original topology. Mathematically, the fault changes the power network's topology and  transforms the power system's evolution from the pre-fault dynamics to fault-on dynamics, which drives away the system from the normal stable operating point to a fault-cleared state at the clearing time, i.e. the time instant at which the fault that disturbed the system is cleared or self-clears. With a set of faults, then we have a set of  fault-cleared states at a given clearing time. The mathematical approach developed in this work bounds the reachability set of the fault-on dynamics, and therefore the set of fault-cleared states. This allows us to certify that these fault-cleared states remain in the attraction region of the original equilibrium point, and thus ensuring that the grid is still stable after suffering the attack of faults. This type of robust resiliency assessment is completely simulation-free, unlike the widely adopted controlling-UEP approaches that rely on simulations of the fault-on dynamics.


The third innovation of this paper is the introduction of the
quadratic Lyapunov functions for transient stability assessment of
power grids. Existing approaches to this problem are based on
energy function \cite{Chiang:2011eo} and Lur'e-Postnikov type
Lyapunov function \cite{Hill:1989:LFL:72068.72082,
Hiskens:1997Lya, VuTuritsyn:2014},  both of which are nonlinear
non-quadratic and generally non-convex functions. The convexity of
quadratic Lyapunov functions enables the real-time construction of
the stability/resiliency certificate and real-time stability
assessment. This is an advancement compared to the energy function
based methods, where computing the critical UEP for stability
analysis is generally an NP-hard problem.

On the computational aspect, it is worthy to note that all the
approaches developed in this work are based on solving
semidefinite programming (SDP) with matrices of sizes smaller than
two times of the number of buses or transmission lines (which
typically scales linearly with the number of buses due to the
sparsity of power networks). For large-scale power systems,
solving these problems with off-the-shelf solvers may be slow.
However, it was shown in a number of recent studies that matrices
appearing in power system context are characterized by graphs with
low maximal clique order. This feature is efficiently exploited in
a new generation of SDP solvers \cite{OPT-006} enabling the
related SDP problems to be quickly solved by SDP relaxation and
decomposition methods. Moreover, an important advantage of the
robust certificates proposed in this work is that they allow the
computationally cumbersome task of calculating the suitable
Lyapunov function and corresponding critical value to be performed
off-line, while the much more cheaper computational task of
checking the stability/resilience condition will be carried out
online. In this manner, the proposed certificates can be used in
an extremely efficient way as a complementary method together with
other direct methods and time domain simulations for contingency
screening, yet allowing for effectively screening of many
non-critical contingencies.

\subsection{Relevant Work}

In \cite{VuTuritsyn:2014}, we introduced the Lyapunov functions family
approach to transient stability of power system. This approach
 can certify stability for a large set of fault-cleared
states, deal with losses in the systems \cite{VuTuritsyn:2014acc},
and is possibly applicable to structure-preserving model and
higher-order models of power grids \cite{VuTuritsyn:2014pes}.
However, the possible non-convexity of Lyapunov functions in
Lur'e-Postnikov form requires to relax this approach to make the
stability certificate scalable to large-scale power grids. The
quadratic Lyapunov functions proposed in this paper totally
overcomes this difficulty. Quadratic Lyapunov functions were also
utilized in \cite{zhao2014design,mallada2014optimal} to analyze
the stability of power systems under load-side controls. This
analysis is possible due to the linear model of power systems
considered in those works. In this paper, we however consider the
power grids that are strongly nonlinear. Among other works, we
note the practically relevant approaches for transient stability
and security analysis based on convex optimizations
\cite{Backhaus:2014} and power network decomposition technique and
Sum of Square programming \cite{Anghel:2013}. Also, the problem of
stability enforcement for power systems attracted much interest
\cite{ortega2005transient, galaz2003energy, shen2000adaptive},
where the passivity-based control approach was employed.

The paper is structured as follows. In Section \ref{sec.model} we
introduce the standard structure-preserving model of power
systems. On top of this model, we formulate in Section
\ref{sec.formulation} two robust  stability and resiliency
problems of power grids, one involves the uncertainty in the
equilibrium points and the other involves the uncertainty in the
sources of faults. In Section \ref{sec.certificates} we introduce
the quadratic Lyapunov functions-based approach to construct the
robust stability/resiliency certificates. Section
\ref{sec.simulations} illustrates the effectiveness of these
certificates through numerical simulations.

\section{Network Model}
\label{sec.model}

A power transmission grid includes generators, loads, and
transmission lines connecting them. A generator has both internal
AC generator bus and load bus. A load only has load bus but no
generator bus. Generators and loads have their own dynamics
interconnected by the nonlinear AC power flows in the transmission
lines. In this paper we consider the standard structure-preserving
model to describe components and dynamics in power systems
\cite{bergen1981structure}. This model naturally incorporates the
dynamics of generators' rotor angle as well as response of load
power output to frequency deviation. Although it does not model
the dynamics of voltages in the system, in comparison to the
classical swing equation with constant impedance loads the
structure of power grids is preserved in this model.

Mathematically, the grid is described by an undirected graph
$\mathcal{A}(\mathcal{N},\mathcal{E}),$ where
$\mathcal{N}=\{1,2,\dots,|\mathcal{N}|\}$ is the set of buses and
$\mathcal{E} \subseteq \mathcal{N} \times \mathcal{N}$ is the set
of transmission lines connecting those buses. Here, $|A|$ denotes
the number of elements in the set $A.$ The sets of generator buses
and load buses are denoted by $\mathcal{G}$ and $\mathcal{L}$ and
labeled as $\{1,..., |\mathcal{G}|\}$ and $\{|\mathcal{G}|+1,...,
|\mathcal{N}|\}.$ We assume that the grid is lossless with
constant voltage magnitudes $V_k, k\in \mathcal{N},$ and the
reactive powers are ignored.


\textbf{Generator buses.} In general, the dynamics of generators
is characterized by its internal voltage phasor. In the context of
transient stability assessment the internal voltage magnitude is
usually assumed to be constant due to its slow variation in
comparison to the angle. As such, the dynamics of the $k^{th}$
generator is described through the dynamics of the internal
voltage angle $\delta_k$ in the so-called swing equation:
\begin{align}
\label{eq.swing1}
  m_k \ddot{\delta_k} + d_k \dot{\delta_k} + P_{e_k}-P_{m_k}
  =0, k \in \mathcal{G},
\end{align}
where, $m_k>0$ is the dimensionless moment of inertia of the
generator, $d_k>0$ is the term representing primary frequency
controller action on the governor, $P_{m_k}$ is the input shaft power producing the mechanical torque acting on the rotor, and $P_{e_k}$
is the effective dimensionless electrical power output of the
$k^{th}$ generator.

\textbf{Load buses.} Let $P_{d_k}$ be the real power drawn by the
load at $k^{th}$ bus, $k \in \mathcal{L}$. In general $P_{d_k}$ is
a nonlinear function of voltage and frequency. For constant
voltages and small frequency variations around the operating point
$P^0_{d_k}$, it is reasonable to assume that
\begin{align}
P_{d_k}=P^0_{d_k} + d_k \dot{\delta}_k, k \in \mathcal{L},
\end{align}
where $d_k>0$ is  the constant frequency coefficient of load.

\textbf{AC power flows.} The active electrical power $P_{e_k}$
injected from the
$k^{th}$ bus into the network, where $k \in \mathcal{N},$ is given by
\begin{align}
\label{eq.electricpower}
  P_{e_k}=\sum_{j \in
  \mathcal{N}_k} V_kV_jB_{kj} \sin(\delta_k
  -\delta_j), k\in \mathcal{N}.
\end{align}
Here, the value $V_k$ represents the voltage magnitude of the
$k^{th}$ bus which is assumed to be constant; $B_{kj}$ is the
(normalized)  susceptance of the transmission line $\{k,j\}$ connecting the $k^{th}$ bus and $j^{th}$ bus;
$\mathcal{N}_k$ is the set of neighboring buses of the $k^{th}$
bus. Let $a_{kj}=V_kV_jB_{kj}.$ By power balancing we obtain the
structure-preserving model of power systems as:
\begin{subequations}
\label{eq.structure-preserving}
\begin{align}
\label{eq.structure-preserving1}
 m_k \ddot{\delta_k} + d_k \dot{\delta_k} + \sum_{j \in
  \mathcal{N}_k} a_{kj} \sin(\delta_k-\delta_j) = &P_{m_k},  k \in \mathcal{G},  \\
  \label{eq.structure-preserving2}
  d_k \dot{\delta_k} + \sum_{j \in
  \mathcal{N}_k} a_{kj} \sin(\delta_k-\delta_j) = &-P^0_{d_k},  k \in \mathcal{L},
\end{align}
\end{subequations}
where, the equations \eqref{eq.structure-preserving1} represent
the dynamics at generator buses and the equations
\eqref{eq.structure-preserving2} the dynamics at  load buses.

The system described by equations \eqref{eq.structure-preserving}
has many stationary points with at least one stable corresponding
to the desired operating point. Mathematically, the state of
\eqref{eq.structure-preserving} is presented by
$\delta=[\delta_1,...,\delta_{|\mathcal{G}|},\dot{\delta}_1,...,\dot{\delta}_{|\mathcal{G}|},\delta_{|\mathcal{G}|+1},...,\delta_{|\mathcal{N}|}]^T,$
and the desired operating point is characterized by the buses'
angles
$\delta^*=[\delta_1^*,...,\delta_{|\mathcal{G}|}^*,0,\dots,0,\delta^*_{|\mathcal{G}|+1},...,\delta^*_{|\mathcal{N}|}]^T.$
This point is not unique since any shift in the buses' angles
$[\delta_1^*+c,...,\delta_{|\mathcal{G}|}^*+c,0,\dots,0,\delta^*_{|\mathcal{G}|+1}+c,...,\delta^*_{|\mathcal{N}|}+c]^T$
is also an equilibrium. However, it is unambiguously characterized
by the angle differences $\delta_{kj}^*=\delta_k^*-\delta_j^*$
that solve the following system of power-flow like equations:
\begin{align}
  \label{eq.SEP}
  \sum_{j \in
  \mathcal{N}_k} a_{kj} \sin(\delta_{kj}^*) =P_{k}, k \in \mathcal{N},
\end{align}
where $P_k=P_{m_k}, k \in \mathcal{G},$ and $P_k=-P^0_{d_k}, k \in
\mathcal{L}.$

\textbf{Assumption:} There is a solution $\delta^*$ of equations
\eqref{eq.SEP} such that $|\delta_{kj}^*| \le \gamma < \pi/2$ for
all the transmission lines $\{k,j\} \in \mathcal{E}.$

We recall that for almost all power systems this assumption holds
true if we have the following synchronization condition, which is
established in \cite{Dorfler:2013},
\begin{align}
\label{eq.SynchronizationCondition}
\|L^{\dag}p\|_{\mathcal{E},\infty} \le \sin\gamma.
\end{align}
Here, $L^\dag$ is the pseudoinverse of the network Laplacian
matrix, $p=[P_1,...,P_{|\mathcal{N}|}]^T,$ and
$\|x\|_{\mathcal{E},\infty}=\max_{\{i,j\}\in
\mathcal{E}}|x(i)-x(j)|.$ In the sequel, we denote as
$\Delta(\gamma)$ the set of equilibrium points $\delta^*$
satisfying  that $|\delta_{kj}^*| \le \gamma<\pi/2, \forall
\{k,j\}\in \mathcal{E}.$ Then, any equilibrium point in this set
is a stable operating point \cite{Dorfler:2013}.

We note that, beside $\delta^*$ there are many other solutions of
\eqref{eq.SEP}. As such, the power system
\eqref{eq.structure-preserving} has many equilibrium points, each
of which has its own region of attraction. Hence, analyzing the
stability region of the stable equilibrium point $\delta^*$ is a
challenge to be addressed in this paper.

\section{Robust Stability and Resiliency Problems}
\label{sec.formulation}

\begin{figure}[t!]
\centering
\includegraphics[width = 3.2in]{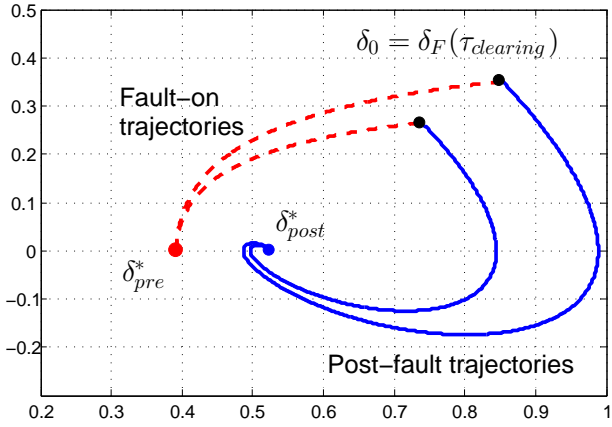}
\caption{Convergence of the post-fault dynamics from two different fault-cleared states $\delta_F(\tau_{clearing}),$
which are obtained from two different fault-on dynamics at the clearing times $\tau_{clearing},$ to the post-fault equilibrium point $\delta^*_{post}.$}
\label{fig.Screening}
\end{figure}
\subsection{Contingency Screening for Transient Stability}
In contingency screening for transient stability, we consider
three types of dynamics of power systems, namely pre-fault
dynamics, fault-on dynamics and post-fault dynamics. In normal
conditions, a power grid operates at a stable equilibrium point of
the pre-fault dynamics. After the initial disturbance,
the system evolves according to the fault-on dynamics laws and moves away
from the pre-fault equilibrium point $\delta^*_{pre}$. After some
time period, the fault is cleared or self-clears, and the system is at the
fault-cleared state $\delta_0=\delta_F(\tau_{clearing})$. Then, the power system experiences the
post-fault transient dynamics.  The
transient stability assessment problem addresses the question of whether the post-fault
dynamics converges from the fault-cleared state to a post-fault
stable equilibrium point $\delta^*_{post}$.
Figure \ref{fig.Screening} shows the transient stability of the post-fault dynamics originated
from the fault-cleared states to the stable post-fault equilibrium.

\subsection{Problem Formulation}
The robust transient stability problem involves situations where there is
uncertainty in power injections $P_k,$ the sources of which are intermittent renewable generations and varying power consumptions.
Particularly, while the
parameters $m_k,d_k$ are fixed and known, the power generations
$P_{m_k}$ and load consumption $P^0_{d_k}$ are changing in time. As such, the post-fault equilibrium
$\delta_{post}^*$ defined by \eqref{eq.SEP} also variates. This raises the need for a robust stability certificate
that can certify stability of post-fault dynamics with respect to a set of
equilibria. When the power injections $P_k$ change in each transient
stability assessment cycle, such a robust stability certificate
can be repeatedly utilized in the ``off-line" certification of system  stability, eliminating the need for assessing stability on a regular basis.
Formally, we consider the following robust stability problem:

\begin{itemize}
\item [(\textbf{P1})] \textbf{Robust stability w.r.t. a set of unknown equilibria:} \emph{Given a fault-cleared state $\delta_0,$ certify the transient stability of the post-fault dynamics described by
\eqref{eq.structure-preserving} with respect to the set of
stable equilibrium points $\Delta(\gamma)$.}
\end{itemize}
We note that though the equilibrium point $\delta^*$ is unknown,
we still can determine if it belongs to the set $\Delta(\gamma)$
by checking if the power injections satisfy the synchronization condition
\eqref{eq.SynchronizationCondition} or not.

The robust resiliency property denotes the ability of power systems to withstand
a set of unknown disturbances and recover to the stable operating conditions. We consider the scenario where the disturbance results in line tripping. Then, it self-clears and the faulted line is reclosed. For simplicity, assume that the steady state power injections $P_k$ are unchanged during the fault-on
dynamics. In that case, the pre-fault and post-fault equilibrium
points  defined by \eqref{eq.SEP} are the same:
$\delta_{pre}^*=\delta_{post}^*=\delta^*$ (this assumption is only for simplicity of presentation, we will discuss the case when $\delta_{pre}^* \neq \delta_{post}^*$). However, we assume that
we don't know which line is tripped/reclosed. Hence, there is a
set of possible fault-on dynamics, and we want to certify if the
 power system can withstand this set of faults and recover to the stable condition $\delta^*$.
Formally, this type of robust resiliency is formulated as
follows.

\begin{itemize}
\item [(\textbf{P2})] \textbf{Robust resiliency w.r.t. a set of faults:} \emph{Given a power system with the pre-fault and post-fault equilibrium point $\delta^* \in \Delta(\gamma),$ certify if the
post-fault dynamics will return from any possible fault-cleared state
$\delta_0$ to the equilibrium point $\delta^*$ regardless of the
fault-on dynamics.}
\end{itemize}

To resolve these problems in the next section, we utilize tools
from nonlinear control theory. For this end, we separate the
nonlinear couplings and the linear terminal system in
\eqref{eq.structure-preserving}. For brevity, we denote the stable
post-fault equilibrium point for which we want to certify
stability as $\delta^*.$ Consider the state vector $x =
[x_1,x_2,x_3]^T,$ which is composed of the vector of generator's
angle deviations from equilibrium $x_1 = [\delta_1 -
\delta_1^*,\dots, \delta_{|\mathcal{G}|} -
\delta_{|\mathcal{G}|}^*]^T$, their angular velocities $x_2 =
[\dot\delta_1,\dots,\dot\delta_{|\mathcal{G}|}]^T$, and vector of
load buses' angle deviation from equilibrium
$x_3=[\delta_{{|\mathcal{G}|}+1}-\delta_{{|\mathcal{G}|}+1}^*,\dots,\delta_{|\mathcal{N}|}-\delta_{|\mathcal{N}|}^*]^T$.
Let $E$ be the incidence matrix of the  graph
$\mathcal{G}(\mathcal{N},\mathcal{E})$, so that
$E[\delta_1,\dots,\delta_{|\mathcal{N}|}]^T =
[(\delta_k-\delta_j)_{\{k,j\}\in\mathcal{E}}]^T$. Let the matrix
$C$ be $E[I_{m\times m} \;O_{m \times n};O_{(n-m) \times 2m} \;
I_{(n-m)\times (n-m)}].$ Then
$$Cx=E[\delta_1-\delta_1^*,\dots,\delta_{|\mathcal{N}|}-\delta_{|\mathcal{N}|}^*]^T=[(\delta_{kj}-\delta_{kj}^*)_{\{k,j\}\in\mathcal{E}}]^T.$$

Consider the vector of nonlinear interactions $F$ in the simple trigonometric form: $
F(Cx)=[(\sin\delta_{kj}-\sin\delta^*_{kj})_{\{k,j\}\in\mathcal{E}}]^T.$ Denote
the matrices of moment of inertia, frequency controller  action on governor, and  frequency coefficient of load
as $M_1=\emph{\emph{diag}}(m_1,\dots,m_{|\mathcal{G}|}), D_1=\emph{\emph{diag}}(d_1,\dots,d_{|\mathcal{G}|})$ and $M=\emph{\emph{diag}}(m_1,\dots,m_{|\mathcal{G}|},d_{|\mathcal{G}|+1},\dots,d_{|\mathcal{N}|}).$

In state space representation, the power system \eqref{eq.structure-preserving} can be then expressed in the
following compact form:
\begin{align}
\dot{x}_1 &=x_2 \nonumber \\
\dot{x}_2 &=M_1^{-1}D_1x_2-S_1M^{-1}E^TSF(Cx)  \\
\dot{x}_3 &= -S_2M^{-1}E^TS F(Cx) \nonumber
\end{align}
where $S=\emph{\emph{diag}}(a_{kj})_{\{k,j\}\in \mathcal{E}},
S_1=[I_{m\times m}\quad O_{m\times n-m}], S_2=[O_{n-m\times m} \quad I_{n-m\times n-m}], n=|\mathcal{N}|, m=|\mathcal{G}|.$
Equivalently, we have
\begin{equation}\label{eq.Bilinear}
 \dot x = A x - B F(C x),
\end{equation}
with the matrices $A,B$ given by the following expression:
\begin{align*}
A=\left[
        \begin{array}{ccccc}
          O_{m \times m} \qquad & I_{m \times m}  \qquad & O_{m \times n-m}\\
          O_{m \times m} \qquad & -M_1^{-1}D_1 \qquad & O_{m \times n-m} \\
          O_{n-m \times m} \qquad &O_{n-m \times m} \qquad & O_{n-m \times n-m}
        \end{array}
      \right],
\end{align*}
and $$
 B= \left[
        \begin{array}{ccccc}
          O_{m \times |\mathcal{E}|}; \quad
          S_1M^{-1}E^TS; \quad
          S_2M^{-1}E^TS
        \end{array}
      \right].$$

The key advantage of this state space representation of the system
is the clear separation of nonlinear terms that are represented as
a ``diagonal'' vector function composed of simple univariate
functions applied to individual vector components. This feature
will be exploited to construct Lyapunov functions for stability
certificates in the next section.

\section{Quadratic Lyapunov Function-based Stability and Resiliency Certificates}
\label{sec.certificates}

This section introduces the robust stability and resiliency
certificates to address the problems $\textbf{(P1)}$ and
$\textbf{(P2)}$ by utilizing quadratic Lyapunov functions. The
construction of these quadratic Lyapunov functions is based on
exploiting the strict bounds of the nonlinear vector $F$ in a
region surrounding the equilibrium point and solving a linear
matrix inequality (LMI). In comparison to the typically non-convex
energy functions and Lur'e-Postnikov type Lyapunov functions, the
convexity of quadratic Lyapunov functions enables the quick
construction of the stability/resiliency certificates and the
real-time stability assessment. Moreover, the certificates
constructed in this work rely on the semi-local bounds of the
nonlinear terms, which ensure that nonlinearity $F$ is linearly
bounded in a polytope surrounding the equilibrium point.
Therefore, though similar to the circle criterion, these stability
certificates constitute an advancement to the classical circle
criterion for stability in control theory where the nonlinearity
is linearly bounded in the whole state space.

\subsection{Strict Bounds for Nonlinear Couplings}

\begin{figure}[t!]
\centering
\includegraphics[width = 3.2in]{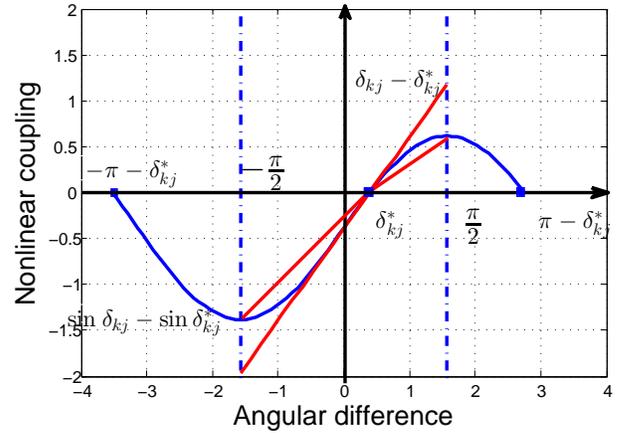}
\caption{Strict bounds of nonlinear sinusoidal couplings $(\sin\delta_{kj}-\sin\delta^*_{kj})$ by two
linear functions of the angular difference $\delta_{kj}$ as described in \eqref{eq.bound}}
\label{fig.NonlinearityBounding}
\end{figure}

The representation \eqref{eq.Bilinear} of the
structure-preserving model \eqref{eq.structure-preserving} with
separation of nonlinear interactions allows us to
naturally bound the nonlinearity of the system in the spirit of
traditional approaches to nonlinear control
\cite{Popov:1962,Yakubovich:1967,Megretski:1997}.  Indeed, Figure
\ref{fig.NonlinearityBounding} shows the natural bound of the
nonlinear interactions $(\sin\delta_{kj} - \sin\delta_{kj}^*)$ by the linear functions
of angular difference $(\delta_{kj} - \delta_{kj}^*)$. From Fig.
\ref{fig.NonlinearityBounding}, we observe that for all values of
$\delta_{kj} = \delta_k - \delta_j$  such that $|\delta_{kj}|  \le
\pi/2,$ we have:
\begin{align}
 g_{kj}(\delta_{kj}-\delta_{kj}^*)^2 \le (\delta_{kj}-\delta_{kj}^*)(\sin\delta_{kj} - \sin\delta_{kj}^*) \le (\delta_{kj}-\delta_{kj}^*)^2
\end{align}
where
\begin{align}
g_{kj}=\min \{\frac{1-\sin\delta_{kj}^*}{\pi/2-\delta_{kj}^*},
\frac{1+\sin\delta_{kj}^*}{\pi/2+\delta_{kj}^*}\} =
\frac{1-\sin|\delta_{kj}^*|}{\pi/2-|\delta_{kj}^*|}
\end{align}
As the function $(1-\sin t)/(\pi/2-t)$ is decreasing on
$[0,\pi/2],$ it holds that
\begin{align}
\label{eq.gain} g_{kj} \ge \frac{1-\sin \lambda(\delta^*)}{\pi/2-
\lambda(\delta^*)}:=g>0
\end{align}
where $\lambda(\delta^*)$ is the maximum value of
$|\delta_{kj}^*|$ over all the lines $\{k,j\} \in \mathcal{E},$
and $0\le \lambda(\delta^*)\le \gamma <\pi/2.$ Therefore, in the
polytope $\mathcal{P},$ defined by inequalities $|\delta_{kj}| \le
\pi/2,$ all the elements of the nonlinearities $F$ are bounded by:
\begin{align}
\label{eq.bound} g(\delta_{kj}-\delta_{kj}^*)^2 \le
(\delta_{kj}-\delta_{kj}^*)(\sin\delta_{kj} - \sin\delta_{kj}^*)
\le (\delta_{kj}-\delta_{kj}^*)^2
\end{align}
and hence,
\begin{align}
\big(F(Cx)-gCx \big)^T\big(F(Cx) -Cx\big) \le 0, \forall x\in \mathcal{P}.
\end{align}

\subsection{Quadratic Lyapunov Functions}
In this section, we  introduce the quadratic Lyapunov functions to
analyze the stability of the general Lur'e-type system
\eqref{eq.Bilinear}, which will be instrumental to the
constructions of stability and resiliency certificates in this paper.
The certificate construction is based on the following result which can be seen as an extension of the classical circle
criterion to the case when the sector bound condition only holds in a finite region.

\begin{lemma}
\label{thr.QuadraticLyapunov} \emph{Consider the general system in
the form \eqref{eq.Bilinear} in which the nonlinear
vector $F$ satisfies the sector bound condition that $(F-K_1Cx)^T(F-K_2Cx) \le 0$ for some
matrices $K_1,K_2$ and $x$ belonging to the set $\mathcal{S}.$
Assume that there exists a positive definite matrix $P$ such that
\begin{align}
\label{eq.ConditionP} A^TP +PA- C^TK_1^TK_2C + R^TR \le 0,
\end{align}
where $R=B^TP-\frac{1}{2}(K_1+K_2)C.$ Then, the quadratic Lyapunov
function $V(x(t))=x(t)^TPx(t)$ is decreasing along trajectory of
the system \eqref{eq.Bilinear} whenever $x(t)$ is in the set
$\mathcal{S}$.}
\end{lemma}

\emph{Proof:} See Appendix \ref{appendix.QuadraticLyapunov}. $\square$

Note that when $K_1=0$ or $K_2=0$ then Condition
\eqref{eq.ConditionP} leads to that the matrix $A$ have to be
strictly stable. This condition does not hold for the case of
structure-preserving model \eqref{eq.structure-preserving}. Hence in case when
$K_1=0$ or $K_2=0$, it is hard to have a quadratic
Lyapunov function certifying the convergence of the system \eqref{eq.structure-preserving}  by Lemma \ref{thr.QuadraticLyapunov}.
Fortunately, when we restrict the system state $x$ inside the polytope
$\mathcal{P}$ defined by inequalities $|\delta_{kj}| \le \pi/2,$
we have strict bounds for the nonlinear interactions $F$ as in
\eqref{eq.bound}, in which $K_1=gI, K_2=I$ are strictly positive.
Therefore, we can obtain the quadratic Lyapunov function certifying convergence of the
structure-preserving model \eqref{eq.structure-preserving} as
follows.

\begin{lemma}
\label{thr.Lyapunov} \emph{Consider power grids described by the
structure-preserving model \eqref{eq.structure-preserving} and
satisfying Assumption 1. Assume that for given matrices
$A,B,C,$ there exists a positive definite matrix $P$ of size $(|\mathcal{N}|+|\mathcal{G}|)$ such that
\begin{align}
&(A-\frac{1}{2}(1+g)BC)^TP+P(A-\frac{1}{2}(1+g)BC) \nonumber \\&+
PBB^TP + \frac{(1-g)^2}{4}C^TC \le  0
\end{align}
or equivalently (by Schur complement) satisfying the LMI
\begin{align}
\label{eq.KeyCondition}
\left[%
\begin{array}{cc}
 \bar{A}^TP+P\bar{A} +  \dfrac{(1-g)^2}{4}C^TC     & PB \\
 B^TP  & -I \\
\end{array}%
\right] \le 0
\end{align}
where $\bar{A}=A-\dfrac{1}{2}(1+g)BC.$ Then, along
\eqref{eq.structure-preserving}, the Lyapunov function $V(x(t))$
is decreasing whenever $x(t) \in \mathcal{P}.$}
\end{lemma}

\emph{Proof:} From \eqref{eq.bound}, we can see that the
 vector of nonlinear interactions $F$ satisfies the sector bound condition: $(F-K_1Cx)^T(F-K_2Cx) \le 0,$
in which $K_1=gI,K_2=I$ and the set $\mathcal{S}$ is the polytope
$\mathcal{P}$ defined by inequalities $|\delta_{kj}| \le \pi/2.$ Applying Lemma
\ref{thr.QuadraticLyapunov}, we have Lemma \ref{thr.Lyapunov}
straightforwardly. $\square$

We observe that the matrix $P$ obtained by solving the LMI
\eqref{eq.KeyCondition} depends on matrices $A,B,C$ and the gain
$g.$ Matrices $A,B,C$ do not depend on the parameters $P_k$ in the
structure preserving model \eqref{eq.structure-preserving}. Hence,
we have a common triple of matrices $A,B,C$ for all the
equilibrium point $\delta^*$ in the set $\Delta(\gamma).$ Also,
whenever $\delta^*\in \Delta(\gamma)$, we can replace $g$ in
\eqref{eq.gain} by the lower bound of $g$ as $g=\dfrac{1-\sin
\gamma}{\pi/2- \gamma} >0.$ This lower bound also does not depend
on the equilibrium point $\delta^*$ at all. Then, the matrix $P$
is independent of the set $\Delta(\gamma)$ of stable equilibrium
points $\delta^*.$ Therefore, Lemma \ref{thr.Lyapunov} provides us
with a common quadratic Lyapunov function for any post-fault
dynamics with post-fault equilibrium point $\delta^* \in
\Delta(\gamma)$. In the next section, we present the transient
stability certificate based on this quadratic Lyapunov function.

\subsection{Transient Stability Certificate}

Before proceeding to robust stability/resiliency certificates  in the next
sections, we will present the transient stability certificate. We
note that the Lyapunov function $V(x)$ considered  in Lemma  \ref{thr.Lyapunov} is
decreasing whenever the system trajectory evolves inside the polytope $\mathcal{P}.$ Outside $\mathcal{P},$ the Lyapunov function
is possible to increase. In the following,
we will construct inside the polytope $\mathcal{P}$ an invariant set $\mathcal{R}$ of the post-fault dynamics described by structure-preserving system
\eqref{eq.structure-preserving}. Then, from
any point inside this invariant set $\mathcal{R}$, the post-fault dynamics
\eqref{eq.structure-preserving} will only evolve inside
$\mathcal{R}$ and eventually converge to the equilibrium point due to the
decrease of the Lyapunov function $V(x)$.

Indeed, for each edge $\{k,j\}$ connecting the generator buses $k$ and $j,$ we divide the boundary $\partial\mathcal{P}_{kj}$ of
$\mathcal{P}$ corresponding to the equality $|\delta_{kj}|=\pi/2$
into two subsets $\partial\mathcal{P}_{kj}^{in}$ and
$\partial\mathcal{P}_{kj}^{out}$. The flow-in boundary segment
$\partial\mathcal{P}_{kj}^{in}$ is defined by
$|\delta_{kj}|=\pi/2$ and $\delta_{kj}\dot{\delta}_{kj} < 0,$
while the flow-out boundary segment
$\partial\mathcal{P}_{kj}^{out}$ is defined by
$|\delta_{kj}|=\pi/2$ and $\delta_{kj}\dot{\delta}_{kj} \ge 0.$
Since the derivative of $\delta_{kj}^2$ at every points on
$\partial\mathcal{P}_{kj}^{in}$ is negative, the system trajectory
of \eqref{eq.structure-preserving} can only go inside
$\mathcal{P}$ once it meets $\partial\mathcal{P}_{kj}^{in}.$

Define the following minimum value of the Lyapunov function $V(x)$ over the flow-out
boundary $\partial\mathcal{P}^{out}$ as:
\begin{align}\label{eq.Vmin1}
 V_{\min}=\mathop {\min}\limits_{x \in \partial\mathcal{P}^{out}} V(x),
\end{align}
where $\partial\mathcal{P}^{out}$ is the flow-out boundary of the
polytope $\mathcal{P}$ that is the union of
$\partial\mathcal{P}_{kj}^{out}$ over all the transmission lines $\{k,j\}\in \mathcal{E}$ connecting generator buses.  From the decrease of $V(x)$
inside the polytope $\mathcal{P},$ we can have the following
center result regarding transient stability assessment.

\begin{theorem}
\label{thr.StabilityAssessment}
  \emph{For a post-fault equilibrium point $\delta^* \in \Delta(\gamma)$, from any initial state $x_0$ staying in set $\mathcal{R}$ defined by
  \begin{align}
  \label{eq.StabilityRegion}
  \mathcal{R}=\{x\in \mathcal{P}: V(x) < V_{\min}\},
  \end{align}
  then, the system trajectory of \eqref{eq.structure-preserving} will only evolve in the set $\mathcal{R}$
  and eventually converge to the stable equilibrium point $\delta^*.$}
\end{theorem}

\emph{Proof:}
See Appendix \ref{appen.StabilityAssessment}. $\square$

\begin{remark}
Since the Lyapunov function $V(x)$ is convex, finding the minimum
value $V_{\min}=\min_{x \in \partial\mathcal{P}^{out}} V(x)$ can
be extremely fast. Actually, we can have analytical form of
$V_{\min}.$ This fact together with the LMI-based construction of the Lyapunov function $V(x)$
allows us to perform the transient stability assessment in the real time.
\end{remark}

\begin{remark}
Theorem \ref{thr.StabilityAssessment} provides a certificate to
determine if the post-fault dynamics will evolve from the
fault-cleared state $x_0$ to the equilibrium point. By this
certificate,  if $x_0\in \mathcal{R},$ i.e. if $x_0 \in
\mathcal{P}$ and $V(x_0)<V_{\min}$, then we are sure that the
post-fault dynamics is stable. If this is not true, then there is
no conclusion for the stability or instability of the post-fault
dynamics by this certificate.
\end{remark}

\begin{remark}
\label{remark.PracticalConfiguration} The transient stability
certificate in Theorem \ref{thr.StabilityAssessment} is effective
to assess the transient stability of post-fault dynamics where the
fault-cleared state is inside the polytope $\mathcal{P}.$ It can
be observed that the polytope $\mathcal{P}$ contains almost all
practically interesting configurations. In real power grids, high
differences in voltage phasor angles typically result in
triggering of protective relay equipment and make the dynamics of
the system more complicated. Contingencies that trigger those
events are rare but potentially extremely dangerous. They should
be analyzed individually with more detailed and realistic models
via time-domain simulations.

\end{remark}

\begin{remark}
The stability certificate in Theorem \ref{thr.StabilityAssessment}
is constructed similarly to that in \cite{VuTuritsyn:2014}. The main
feature distinguishing the certificate in Theorem
\ref{thr.StabilityAssessment} is that it is based on the quadratic
Lyapunov function, instead of the Lur'e-Postnikov type Lyapunov
function as in  \cite{VuTuritsyn:2014}. As such, we can have an analytical
form for $V_{\min}$ rather than determining it by a potentially
non-convex optimization as in \cite{VuTuritsyn:2014}.
\end{remark}

\subsection{Robust Stability w.r.t. Power Injection Variations}
\begin{figure}[t!]
\centering
\includegraphics[width = 3.2in]{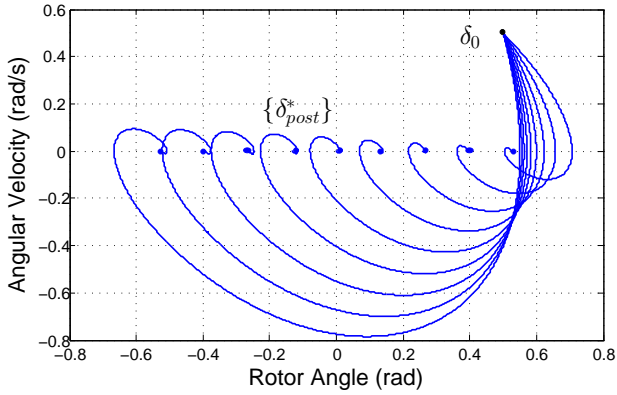}
\caption{Robust transient stability of the post-fault dynamics originated from the fault-cleared state $\delta_0=[0.5 \;\; 0.5]^T$
to the set of stable equilibrium points $\Delta(\pi/6)=\{\delta_{post}^*=[\delta^*,0]^T:-\pi/6\le\delta^*\le\pi/6\}.$}
\label{fig.RobustEquilibria}
\end{figure}

In this section, we develop a ``robust'' extension of the
stability certificate in Theorem \ref{thr.StabilityAssessment}
that can be used to assess transient stability of the post-fault
dynamics described by the structure-preserving model
\eqref{eq.structure-preserving} in the presence of power injection
variations. Specifically, we consider the system whose stable
equilibrium point variates but belongs to the set
$\Delta(\gamma).$ As such, whenever the power injections $P_k$
satisfy the synchronization condition
\eqref{eq.SynchronizationCondition}, we can apply this robust
stability certificate without exactly knowing the equilibrium
point of the system \eqref{eq.structure-preserving}.

As discussed in Remark \ref{remark.PracticalConfiguration}, we are only interested in the case when the fault-cleared
state is in the polytope $\mathcal{P}.$ Denote
$\delta=[\delta_1,...,\delta_{|\mathcal{G}|},\dot{\delta}_1,...,\dot{\delta}_{|\mathcal{G}|},\delta_{|\mathcal{G}|+1},...,\delta_{|\mathcal{N}|}].$
The system state $x$ and the fault-cleared state $x_0$ can be then
presented as $x=\delta-\delta^*$ and $x_0=\delta_0-\delta^*.$ Exploiting the independence of the LMI \eqref{eq.KeyCondition}
on the equilibrium point $\delta^*,$ we
have the following robust stability certificate for the problem $\textbf{(P1)}$.

\begin{theorem}
\label{thr.RobustStability1} \emph{Consider the post-fault dynamics
\eqref{eq.structure-preserving} with uncertain stable equilibrium point
$\delta^*$ that satisfies $\delta^*\in \Delta(\gamma).$ Consider
a fault-cleared state  $\delta_0 \in \mathcal{P},$ i.e., $|\delta_{0_{kj}}|\le\pi/2, \forall \{k,j\}\in \mathcal{E}.$ Suppose that
there exists a positive definite matrix $P$ of size $(|\mathcal{N}|+|\mathcal{G}|)$ satisfying the LMI \eqref{eq.KeyCondition}
and
\begin{align}
\label{eq.RobustCondition}
  \delta_0^TP\delta_0 < \min_{\delta\in \partial\mathcal{P}^{out},
  \delta^*\in\Delta(\gamma)}\big(
  \delta^TP\delta-2 {\delta^*}^TP(\delta-\delta_0)\big)
\end{align}
Then, the system \eqref{eq.structure-preserving} will converge
from the fault-cleared state $\delta_0$ to the equilibrium point $\delta^*$
for any $\delta^*\in \Delta(\gamma).$}
\end{theorem}

\emph{Proof:} See Appendix \ref{appen.RobustStability1}. $\square$

\begin{remark}
Theorem \ref{thr.RobustStability1} gives us a robust certificate to assess the transient stability of the post-fault dynamics \eqref{eq.structure-preserving}
in which the power injections $P_k$ variates. First, we check the synchronization condition \eqref{eq.SynchronizationCondition}, the satisfaction of which
tells us that the equilibrium point $\delta^*$ is in the set $\Delta(\gamma).$ Second, we calculate the positive definite matrix $P$ by solving the LMI \eqref{eq.KeyCondition} where
the gain $g$ is defined as $(1-\sin \gamma)/(\pi/2-\gamma).$ Lastly, for a given fault-cleared state $\delta_0$ staying inside the polytope $\mathcal{P},$ we check
whether the inequality \eqref{eq.RobustCondition} is satisfied or not. In the former case, we conclude that the post-fault dynamics \eqref{eq.structure-preserving} will converge
from the fault-cleared state $\delta_0$ to the equilibrium point $\delta^*$ regardless of the variations in power injections.
Otherwise, we repeat the second step to find other positive definite matrix $P$ and check the condition
\eqref{eq.RobustCondition} again.
\end{remark}

\begin{remark}
Note that there are possibly many matrices $P$ satisfying the LMI
\eqref{eq.KeyCondition}. This gives us flexibility in choosing $P$
satisfying both \eqref{eq.KeyCondition} and
\eqref{eq.RobustCondition} for a given fault-cleared state $\delta_0$.
A heuristic algorithm as in \cite{VuTuritsyn:2014} can be used to find the best suitable matrix $P$ in the
family of such matrices defined by \eqref{eq.KeyCondition} for the given fault-cleared state $\delta_0$ after a finite number of steps.
\end{remark}

\begin{remark}
In practice, to reduce the conservativeness and computational time
in the assessment process, we can off-line compute the common
matrix $P$ for any equilibrium point $\delta^* \in \Delta(\gamma)$
and check on-line the condition $V(x_0) < V_{\min}$ with the data
(initial state $x_0$ and power injections $P_k$) obtained on-line.
In some case the initial state can be predicted before hand, and
if there exists a positive definite matrix $P$ satisfying the LMI
\eqref{eq.KeyCondition} and the inequality
\eqref{eq.RobustCondition}, then the on-line assessment is reduced
to just checking condition \eqref{eq.SynchronizationCondition} for
the power injections $P_k.$
\end{remark}


\subsection{Robust Resiliency w.r.t. a Set of Faults}
In this section, we introduce the robust resiliency
certificate  with respect to a set of faults to solve the problem $\textbf{(P2)}$. We consider the case
when the fault results in tripping of a line. Then it self-clears and the line is reclosed. But we don't know which
line is tripped/reclosed. Note that the pre-fault equilibrium and post-fault
equilibrium, which are obtained by solving the power flow equations \eqref{eq.SEP}, are the same and given.

With the considered set of faults, we have a set of
corresponding fault-on dynamic flows, which drive the system from the
pre-fault equilibrium point to a set of fault-cleared states at
the clearing time. We will
introduce technique to bound the fault-on dynamics, by which we
can bound the set of reachable fault-cleared states. With this way,
we make sure that the reachable set of fault-cleared states remain in the region of attraction of the post-fault equilibrium point, and thus the
post-fault dynamics is stable.

Indeed, we first introduce the resiliency certificate for one fault associating with one faulted transmission line, and then extend it to the robust
resiliency certificate for any faulted line. With the fault of tripping the transmission line
$\{u,v\} \in\mathcal{E}$, the corresponding fault-on dynamics can be obtained from the structure-preserving model \eqref{eq.structure-preserving}
after eliminating the nonlinear interaction $a_{uv}\sin\delta_{uv}.$ Formally, the fault-on dynamics is described by
\begin{align}
\label{eq.Faulton}
\dot{x}_F=Ax_F-BF(Cx_F) +B D_{\{u,v\}}\sin\delta_{F_{uv}},
\end{align}
where $D_{\{u,v\}}$ is the unit vector to extract the $\{u,v\}$ element
from the vector of nonlinear interactions $F.$ Here, we denote the fault-on trajectory as $x_F(t)$
to differentiate it from the post-fault trajectory $x(t).$
We have the following  resiliency certificate for the power system
with equilibrium point $\delta^*$ subject to the faulted-line $\{u,v\}$ in the set $\mathcal{E}$.
\begin{figure}[t!]
\centering
\includegraphics[width = 3.2in]{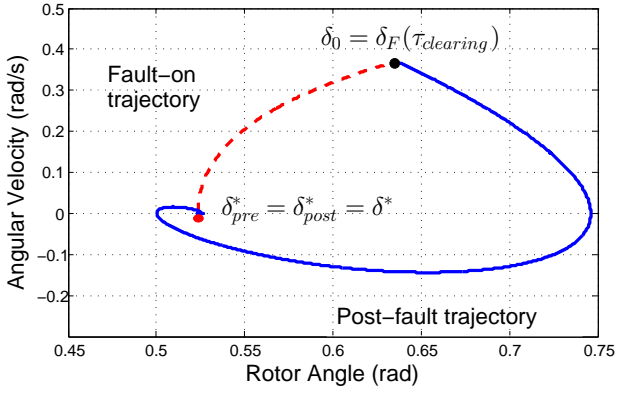}
\caption{Robust resiliency of power system with respect to the faults whenever
the clearing time $\tau_{clearing}<\mu V_{\min}.$}
\label{fig.RobustFault}
\end{figure}
\begin{theorem}
\label{thr.RobustClearingTimeCertificate} \emph{Assume that there exist a positive definite
matrix $P$ of size $(|\mathcal{N}|+|\mathcal{G}|)$ and a positive number $\mu$ such that
\begin{align}
\label{eq.RobustCondition2}
& \bar{A}^TP+P\bar{A} +  \dfrac{(1-g)^2}{4}C^TC  \nonumber \\& + PBB^TP +\mu PBD_{\{u,v\}}D_{\{u,v\}}^TB^TP  \le 0.
\end{align}
Assume that the clearing time $\tau_{clearing}$ satisfies
$\tau_{clearing}< \mu V_{\min}$ where $V_{\min}=\min_{x \in
\partial\mathcal{P}^{out}} V(x)$. Then,
the
fault-cleared state $x_F(\tau_{clearing})$ resulted from the fault-on dynamics \eqref{eq.Faulton}  is still inside the
region of attraction of the post-fault equilibrium point $\delta^*$, and the
post-fault dynamics following the tripping and reclosing of the line $\{u,v\}$ returns to
the original stable operating condition.}
\end{theorem}

\emph{Proof:} See Appendix
\ref{appen.RobustClearingTimeCertificate}.

\begin{remark}
Note that the inequality \eqref{eq.RobustCondition2} can be rewritten as
\begin{align}
\label{eq.RobustCondition3}
 \bar{A}^TP+P\bar{A} +  \dfrac{(1-g)^2}{4}C^TC  + P\bar{B}\bar{B}^TP  \le 0,
\end{align}where $\bar{B}=[B \; \; \sqrt{\mu}BD_{\{u,v\}}].$ By Schur complement, inequality \eqref{eq.RobustCondition3} is equivalent with
\begin{align}
\label{eq.RobustCondition4}
\left[%
\begin{array}{cc}
 \bar{A}^TP+P\bar{A} +  \dfrac{(1-g)^2}{4}C^TC     & P\bar{B} \\
 \bar{B}^TP  & -I \\
\end{array}%
\right] \le 0.
\end{align}
With a fixed value of $\mu$, the inequality \eqref{eq.RobustCondition4} is an LMI which can be transformed to a convex optimization problem.
As such, the inequality \eqref{eq.RobustCondition2} can be solved quickly by a heuristic algorithm in which we vary $\mu$ and find $P$ accordingly from the LMI \eqref{eq.RobustCondition4}
with fixed $\mu$. Another heuristic algorithm to solve the inequality \eqref{eq.RobustCondition2} is to solve the LMI \eqref{eq.KeyCondition}, and for each solution $P$ in this family of solutions, find the maximum value of $\mu$ such that \eqref{eq.RobustCondition2} is satisfied.
\end{remark}

\begin{remark}
For the case when the pre-fault and post-fault equilibrium points
are different, Theorem \ref{thr.RobustClearingTimeCertificate}
still holds true if we replace the condition $\tau_{clearing}< \mu
V_{\min}$ by condition $\tau_{clearing}< \mu
\big(V_{\min}-V(x_{pre})\big),$ where
$x_{pre}=\delta^*_{pre}-\delta^*_{post}.$
\end{remark}

\begin{remark}
The resiliency certificate in Theorem
\ref{thr.RobustClearingTimeCertificate} is straightforward to
extend to a robust resiliency certificate with respect to the set
of faults causing tripping and reclosing of transmission lines in
the grids. Indeed, we will find the positive definite matrix $P$
and positive number $\mu$ such that the inequality
\eqref{eq.RobustCondition2} is satisfied for all the matrices
$D_{\{u,v\}}$ corresponding to the faulted line $\{u,v\} \in
\mathcal{E}.$ Let $D$ be a matrix larger than or equals to the
matrices $D_{\{u,v\}}D_{\{u,v\}}^T$ for all the transmission lines
in $\mathcal{E}$ (here, that $X$ is larger than or equals to $Y$
means that $X-Y$ is positive semidefinite). Then, any positive
definite matrix $P$ and positive number $\mu$ satisfying the
inequality \eqref{eq.RobustCondition2}, in which the matrix
$D_{\{u,v\}}D_{\{u,v\}}^T$ is replaced by $D,$ will give us a
quadratic Lyapunov function-based robust stability certificate
with respect to the set of faults similar to Theorem
\ref{thr.RobustClearingTimeCertificate}. Since
$D_{\{u,v\}}D_{\{u,v\}}^T= \emph{\emph{diag}}(0,\dots,1,\dots,0)$
are orthogonal unit matrices, we can see that the probably best
matrix we can have is
$D=\sum_{\{u,v\}\in\mathcal{E}}D_{\{u,v\}}D_{\{u,v\}}^T=I_{|\mathcal{E}|\times
|\mathcal{E}|}.$ Accordingly, we have the following robust
resilience certificate for any faulted line happening in the
system.
\end{remark}

\begin{theorem}
\label{thr.RobustClearingTimeCertificate2} \emph{Assume that there exist a positive definite
matrix $P$ of size $(|\mathcal{N}|+|\mathcal{G}|)$ and a positive number $\mu$ such that
\begin{align}
\label{eq.RobustCondition5}
& \bar{A}^TP+P\bar{A} +  \dfrac{(1-g)^2}{4}C^TC  + (1+\mu)PBB^TP   \le 0.
\end{align}
Assume that the clearing time $\tau_{clearing}$ satisfies
$\tau_{clearing}< \mu V_{\min}$ where $V_{\min}=\min_{x \in
\partial\mathcal{P}^{out}} V(x)$. Then, for any faulted line happening in the system
the
fault-cleared state $x_F(\tau_{clearing})$ is still inside the
region of attraction of the post-fault equilibrium point $\delta^*$, and the
post-fault dynamics returns to
the original stable operating condition regardless of the fault-on dynamics.}
\end{theorem}

\begin{remark}
By the robust resiliency certificate in Theorem
\ref{thr.RobustClearingTimeCertificate2}, we can certify stability
of power  system with respect to any faulted line happens in the
system. This certificate as well as the certificate in Theorem
\ref{thr.RobustClearingTimeCertificate} totally eliminates the
needs for simulations of the fault-on dynamics, which is currently
indispensable in any existing contingency screening methods for
transient stability.
\end{remark}

\section{Numerical Illustrations}
\label{sec.simulations}
\subsection{2-Bus System}

\begin{figure}[t!]
\centering
\includegraphics[width = 3.2in]{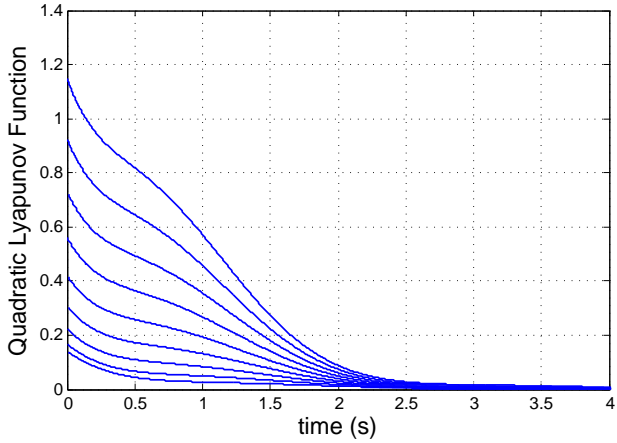}
\caption{Convergence of the quadratic Lyapunov function $V(x)=x^TPx=(\delta-\delta^*)^TP(\delta-\delta^*)$ from the initial value to $0$
when the equilibrium point $\delta^*$ varies in the set $\Delta(\pi/6)=\{\delta_{post}^*=[\delta^*,0]^T:-\pi/6\le\delta^*\le\pi/6\}.$}
\label{fig.RobustEquilibria_Lyapunov}
\end{figure}

For illustration purpose, this section presents the simulation results on the most simple 2-bus power system, described by the single 2-nd
order differential equation
\begin{align}
  m \ddot{\delta} +d \dot{\delta} + a \sin\delta - p=0.
\end{align}
For numerical simulations, we choose $m=0.1$ p.u., $d=0.15$ p.u.,
$a= 0.2$ p.u. When the parameters $p$ changes from $-0.1$ p.u. to
$0.1$ p.u., the stable equilibrium point $\delta^*$ (i.e.
$[\delta^* \; 0]^T$) of the system belongs to the set:
$\Delta=\{\delta^*: |\delta^*| \le \arcsin (0.1/0.2)=\pi/6\}.$ For
the given fault-cleared state $\delta_0=[0.5 \;\;0.5],$ using the
CVX software we obtain a positive matrix $P$ satisfying the LMI
\eqref{eq.KeyCondition} and the condition for robust stability
\eqref{eq.RobustCondition} as $ P= [0.8228  \;\;  0.1402;
    0.1402  \;\;  0.5797].$ The simulations confirm this result. We can see in Fig. \ref{fig.RobustEquilibria} that from the fault-cleared state $\delta_0$
    the post-fault trajectory always converges to the equilibrium point $\delta^*$ for all $\delta^* \in \Delta(\pi/6).$ Figure \ref{fig.RobustEquilibria_Lyapunov}
    shows the convergence of the quadratic Lyapunov function to $0.$

\begin{figure}[t!]
\centering
\includegraphics[width = 3.2in]{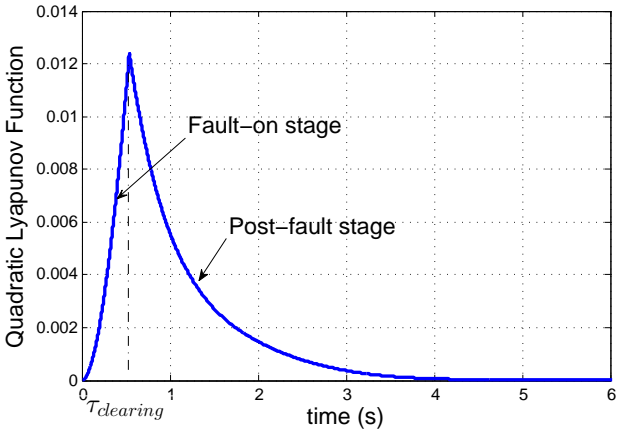}
\caption{Variations of the quadratic Lyapunov function $V(x)=x^TPx=(\delta-\delta^*)^TP(\delta-\delta^*)$ during the fault-on and post-fault dynamics.}
\label{fig.RobustFault_Lyapunov}
\end{figure}

Now we consider the resiliency certificate in Theorem
\ref{thr.RobustClearingTimeCertificate} with respect to fault of
tripping the line and self-clearing. The pre-fault and post-fault
dynamics have the fixed equilibrium point: $\delta^*=[\pi/6
\;0]^T.$ Then the positive definite matrix $P=[0.0822  \;\;
0.0370;
    0.0370 \;\;   0.0603]$ and positive number $\mu=6$ is a solution of the inequality \eqref{eq.RobustCondition2}. As such, for any clearing time $\tau_{clearing}<\mu V_{min}=0.5406,$
    the fault-cleared state is still in the region of attraction of $\delta^*,$ and the power system withstands the fault. Figure \ref{fig.RobustFault}
    confirms this prediction. Figure \ref{fig.RobustFault_Lyapunov} shows that during the fault-on dynamics, the Lyapunov function
    is strictly increasing. After the clearing time $\tau_{clearing},$ the Lyapunov function decreases to $0$ as the post-fault trajectory
    converges to the equilibrium point $\delta^*.$

\subsection{Robust Resiliency Certificate for 3-Generator System}

To illustrate the effectiveness of the robust resiliency certificate in Theorem \ref{thr.RobustClearingTimeCertificate2}, we consider the system of three generators with the time-invariant terminal voltages and mechanical torques given in Tab. \ref{tab.data}.

\begin{table}[ht!]
\centering
\begin{tabular}{|c|c|c|}
  \hline
  Node & V (p.u.) & P (p.u.) \\
  \hline
  1 & 1.0566 & -0.2464 \\
  2 & 1.0502 & 0.2086 \\
  3 & 1.0170 & 0.0378 \\
  \hline
\end{tabular}
\caption{Voltage and mechanical input} \label{tab.data}
\end{table}

The susceptance of the transmission lines are $B_{12}=0.739$ p.u., $B_{13}=1.0958$ p.u., and
$B_{23}=1.245$ p.u. The equilibrium point is calculated from \eqref{eq.SEP}: $\delta^*=[-0.6634\;
   -0.5046\;
   -0.5640 \;0\;0\;0]^T.$ By using CVX software we can find one solution of the inequality \eqref{eq.RobustCondition5} as
   $\mu=0.3$ and the positive definite matrix $P$ as
 \begin{align*}
\left[%
\begin{array}{cccccc}
  2.4376 &   1.7501 &   1.8190  &  4.0789  &  3.9566  &  3.9780\\
    1.7501  &  2.3991  &  1.8576  &  3.9639 &   4.0710  &  3.9785\\
    1.8190  &  1.8576  &  2.3302  &  3.9707 &   3.9859  &  4.0569\\
    4.0789  &  3.9639  &  3.9707  & 17.2977 &  16.6333  & 16.7452\\
    3.9566  &  4.0710  &  3.9859  & 16.6333 &  17.2425  & 16.8003\\
    3.9780  &  3.9785  &  4.0569  & 16.7452 &    16.8003 &  17.1306 \\
\end{array}%
\right]
 \end{align*}
 The corresponding minimum value of Lyapunov function is $V_{\min}=0.5536.$ Hence, for any faults resulting in tripping and reclosing lines in $\mathcal{E},$
 whenever the clearing time less than $\mu V_{\min}=0.1661,$ then the power system still withstands all the faults and recovers to the stable operating condition at $\delta^*.$

\subsection{118 Bus System}

\begin{figure}[t!]
\centering
\includegraphics[width = 3.2in]{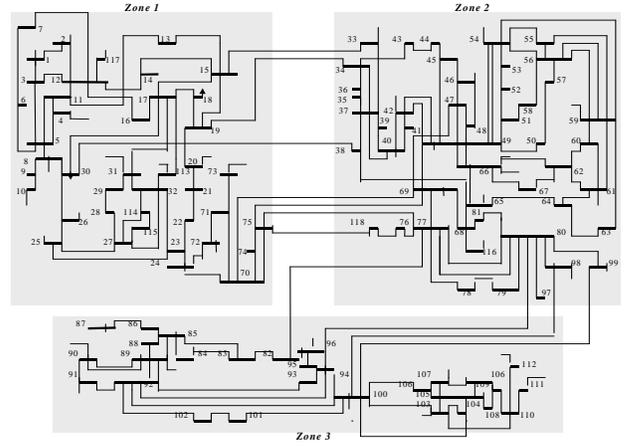}
\caption{IEEE 118-bus test case}
\label{fig.IEEE118}
\end{figure}

Our test system in this section is the  modified IEEE 118-bus test
case \cite{118bus}, of which 54 are generator buses and the other
64 are load buses as showed in Fig. \ref{fig.IEEE118}. The data is
taken directly from the test files \cite{118bus}, otherwise
specified. The damping and inertia are not given in the test files
and thus are randomly selected in the following ranges: $m_i \in
[2,4], \forall i \in \mathcal{G} ,$ and $d_i\in [1,2], \forall i
\in \mathcal{N}.$    The grid originally contains 186 transmission
lines. We eliminate 9 lines whose susceptance is zero, and combine
7 lines $\{42,49\},\{49,54\},\{56,59\},\{49,66\},\{77,80\},$
$\{89,90\},$ and $\{89,92\},$ each of which contains double
transmission lines as in the test files \cite{118bus}.  Hence, the
grid is reduced to 170 transmission lines connecting 118 buses. We
renumber the generator buses as $1-54$ and load buses as $55-118$.

\vskip 0.2cm
\noindent 1) Stability Assessment

\vskip 0.2cm

\begin{figure}[t!]
\centering
\includegraphics[width = 3.2in]{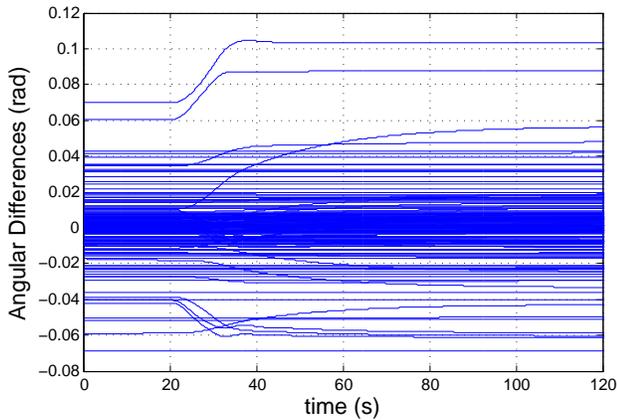}
\caption{Transition of the 118-bus system from the old equilibrium to the new equilibrium when the renewable generations increase $50\%$ in the period $[20s,30s]$}
\label{fig.EquilibriumChange}
\end{figure}

\begin{figure}[t!]
\centering
\includegraphics[width = 3.2in]{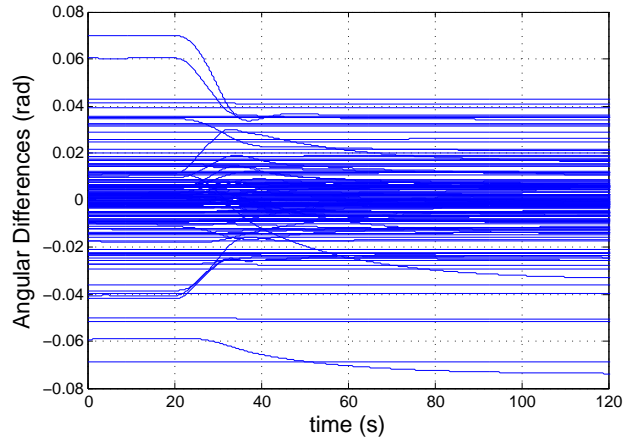}
\caption{Transition of the 118-bus system from the old equilibrium to the new equilibrium when the renewable generations decrease $50\%$ in the period $[20s,30s]$}
\label{fig.EquilibriumChange2}
\end{figure}

We assume that there are varying generations (possibly due to
renewable) at 16 buses $1-16$ (i.e. $30\%$ generator buses are
varying). The system is initially at the equilibrium point given
in \cite{118bus}, but the variations in the renewable generations
make the operating condition to change. We want to assess if the
system will transiently evolve from the initial state to the new
equilibrium points. To make our proposed robust stability
assessment framework valid, we assume that the renewable
generators have the similar dynamics with the conventional
generator but with the varying power output. This happens when we
equip renewable generators with synchronverter
\cite{Zhong_Synchronverters}, which will control the dynamics of
renewables to mimic the dynamics of conventional generators. Using
the CVX software with Mosek solver, we can see that there exists
positive definite matrix $P$ satisfying the LMI
\eqref{eq.KeyCondition} and the inequality
\eqref{eq.RobustCondition} with $\gamma = \pi/12.$ As such, the
grid will transiently evolve from the initial state to any new
equilibrium point in the set $\Delta(\pi/12).$ To demonstrate this
result by simulation, we assume that in the time period
$[20s,30s],$ the power outputs of the renewable generators
increase $50\%.$ Since the synchronization condition
$\|L^{\dag}p\|_{\mathcal{E},\infty} =0.1039 < \sin(\pi/12)$ holds
true, we can conclude that the new equilibrium point, obtained
when the renewable generations increased $50\%,$ will stay in the
set $\Delta(\pi/12).$ From Fig.  \ref{fig.EquilibriumChange}, we
can see that the grid transits from the old equilibrium point to
the new equilibrium point when the renewable power outputs
increase. Similarly, if in the time period $[20s,30s]$ the power
outputs of the renewable generators decrease $50\%,$ then we can
check that $\|L^{\dag}p\|_{\mathcal{E},\infty}=0.0762 <
\sin(\pi/12)$. Therefore by the robust stability certificate, we
conclude that the grid evolves from the old equilibrium point to
the new equilibrium point, as confirmed in Fig.
\ref{fig.EquilibriumChange2}.


\vskip 0.2cm \noindent 2) Resiliency Assessment \vskip 0.2cm We
note that in many cases in practice, when the fault causes
tripping one line, we end up with a new power system with a stable
equilibrium point possibly staying inside the small polytope
$\Delta(\gamma).$ As such, using the robust stability assessment
in the previous section we can certify that, if the fault is
permanent, then the system will transit from the old equilibrium
point to the new equilibrium point. Therefore, to demonstrate the
resiliency certificate we do not need to consider all the tripped
lines, but only concern the case when tripping a critical line may
result in an unstable dynamics, and we use the resiliency
assessment framework to determine if the clearing time is small
enough such that the post-fault dynamics recovers to the old
equilibrium point.

Consider such a critical case when the transmission line
connecting the generator buses 19 and 21 is tripped. It can be
checked that in this case the synchronization condition
\eqref{eq.SynchronizationCondition} is not satisfied even with
$\gamma \approx \pi/2$ since $\|L^{\dag}p\|_{\mathcal{E},\infty}
=1.5963 > \sin(\pi/2).$  As such, we cannot make sure that the
fault-on dynamics caused by tripping the transmission line
$\{19,21\}$ will converge to a stable equilibrium point in the set
$\Delta(\pi/2)$. Now assume that the fault self-clears and the
transmission line $\{19,21\}$ is reclosed at the clearing time
$\tau_{clearing}$. With $\mu=0.11$ and using CVX software with
Mosek solver on a laptop (Intel-core i5 2.6GHz, 8GB RAM), it takes
$1172s$ to find a positive definite matrix $P$ satisfying the
inequality \eqref{eq.RobustCondition4} and to calculate the
corresponding minimum value of Lyapunov function as
$V_{\min}$=0.927. As such, whenever the clearing time satisfies
$\tau_{clearing}<\mu V_{\min}=0.102s$, then the fault-cleared
state is still inside the region of attraction of the post-fault
equilibrium point, i.e. the power system withstands the tripping
of critical line and recover to its stable operating condition.


\section{Conclusions and Path Forward}
This paper has formulated two novel robust stability and
resiliency problems for nonlinear power grids. The first problem
is the transient stability of a given fault-cleared state with
respect to a set of varying post-fault equilibrium points,
particularly applicable to power systems with varying power
injections. The second one is the resiliency of power systems
subject to a set of unknown faults, which result in line tripping
and then self-clearing. These robust stability and resiliency
certificates can help system operators screen multiple
contingencies and multiple power injection profiles, without
relying on computationally wasteful real-time simulations.
Exploiting the strict bounds of nonlinear power flows in a
practically relevant polytope surrounding the equilibrium point,
we introduced the quadratic Lyapunov functions approach to the
constructions of these robust stability/resiliency certificates.
The convexity of quadratic Lyapunov functions allowed us to
perform the stability assessment in the real time.

There are many directions can be pursued to push the introduced
robust stability/resiliency certificates to the industrially ready
level. First, and most important, the algorithms should be
extended to more general higher order models of generators
\cite{Kundur}. Although these models can be expected to be weakly
nonlinear in the vicinity of an equilibrium point, the higher
order model systems are no longer of Lur'e type and have
multi-variate nonlinear terms. It is necessary to extend the
construction from sector-bounded nonlinearities to more general
norm-bounded nonlinearities \cite{boyd1994linear}.

It is also promising to extend the approaches described in this
paper to a number of other problems of high interest to power
system community. These problems include intentional islanding
\cite{januszquiros2014constrained}, where the goal is to identify
the set of tripping signals that can stabilize the otherwise
unstable power system dynamics during cascading failures. This
problem is also interesting in a more general context of designing
and programming of the so-called special protection system that
help to stabilize the system with the control actions produced by
fast power electronics based HVDC lines and FACTS devices.
Finally, the introduced certificates of transient stability can be
naturally incorporated in operational and planning optimization
procedures and eventually help in development of
stability-constrained optimal power flow and unit commitment
approaches \cite{gan2000stability, yuan2003solution}.

\section{Appendix}
\subsection{Proof of Lemma \ref{thr.QuadraticLyapunov}}
\label{appendix.QuadraticLyapunov}

Along the trajectory of \eqref{eq.Bilinear}, we have
\begin{align}
\dot{V}(x) = \dot{x}^TPx + x^TP\dot{x} = x^T(A^TP+PA)x -2 x^TPBF
\end{align}
Let $W(x)=(F-K_1Cx)^T(F-K_2Cx).$ Then, $W(x) \le 0, \forall x \in
\mathcal{S}$ and $W(x)=F^TF-F^T(K_1+K_2)Cx + x^TC^TK_1^TK_2Cx.$
Subtracting $W$ from $\dot{V}(x),$ we obtain:
\begin{align}
&\dot{V}(x) - W(x) = x^T(A^TP+PA)x -2 x^TPBF
\nonumber \\&- F^TF+F^T(K_1+K_2)Cx - x^TC^TK_1^TK_2Cx \nonumber
\\&
               = x^T(A^TP+PA)x -  x^TC^TK_1^TK_2Cx \nonumber \\& - \big \|F+(B^TP-\frac{(K_1+K_2)C}{2})x\big \|^2 \nonumber \\&
               + x^T\big[B^TP-\frac{(K_1+K_2)C}{2}\big]^T\big[B^TP-\frac{(K_1+K_2)C}{2}\big]x \nonumber \\ &
               = x^T\big[A^TP+PA- C^TK_1^TK_2C + R^TR \big]x -S^TS,
\end{align}
where $R=B^TP-\frac{1}{2}(K_1+K_2)C$ and
$S=F+(B^TP-\frac{1}{2}(K_1+K_2)C)x.$

Note that \eqref{eq.ConditionP} is equivalent with the existence
of a non-negative matrix $Q$ such that
\begin{align}
A^TP +PA- C^TK_1^TK_2C + R^TR  =-Q
\end{align}
Therefore:
\begin{align}
\label{eq.dotLyapunov} \dot{V}(x) = W(x) -x^TQx -S^TS \le 0,
\forall x\in \mathcal{S}
\end{align}
As such $V(x(t))$ is decreasing along trajectory $x(t)$ of
\eqref{eq.Bilinear} whenever $x(t)$ is in the set $\mathcal{S}.$ {$\square$}

\subsection{Proof of Theorem \ref{thr.StabilityAssessment}}
\label{appen.StabilityAssessment}

The boundary of the set $\mathcal{R}$ defined as in \eqref{eq.StabilityRegion} is composed of segments
which belong to the boundary of the polytope $\mathcal{P}$ and
segments which belong to the Lyapunov function's sublevel set. Due
to the decrease of $V(x)$ in the polytope $\mathcal{P}$ and the
definition of $V_{\min},$ the system trajectory of
\eqref{eq.structure-preserving} cannot escape the set
$\mathcal{R}$ through the flow-out boundary and the sublevel-set
boundary. Also, once the system trajectory of
\eqref{eq.structure-preserving} meets the flow-in boundary, it
will go back inside $\mathcal{R}.$ Therefore, the system
trajectory of \eqref{eq.structure-preserving} cannot escape
$\mathcal{R},$ i.e. $\mathcal{R}$ is an invariant set of
\eqref{eq.structure-preserving}.

Since $\mathcal{R}$ is a subset of the polytope $\mathcal{P},$ from Lemma \ref{thr.Lyapunov} we have $\dot{V}(x(t)) \le
0$ for all $t\ge 0.$ By LaSalle's Invariance Principle, we
conclude that the system trajectory of
\eqref{eq.structure-preserving} will converge to the set $\{x \in
\mathcal{P}:\dot{V}(x)=0\},$ which together with
\eqref{eq.dotLyapunov} means that the system trajectory of
\eqref{eq.structure-preserving} will converge to the equilibrium
point $\delta^*$ or to some stationary points lying on the
boundary of $\mathcal{P}.$ From the decrease of $V(x)$ in the
polytope $\mathcal{P}$ and the definition of $V_{\min},$ we can
see that the second case cannot happen. Therefore, the system
trajectory will converge to the equilibrium point $\delta^*.$  $\square$

\subsection{Proof of Theorem \ref{thr.RobustStability1}}
\label{appen.RobustStability1}

Since the matrix $P,$ the polytope $\mathcal{P},$ and the fault-cleared state $\delta_0$ are independent of the equilibrium point $\delta^*,$ we have
\begin{align}
V_{\min}-V(x_0) &= \mathop {\min}\limits_{x \in
\partial\mathcal{P}^{out}}\big((\delta-\delta^*)^TP(\delta-\delta^*)\nonumber \\& \qquad\qquad\quad-(\delta_0-\delta^*)^TP(\delta_0-\delta^*)\big) \nonumber \\& =\mathop
{\min}\limits_{\delta \in
\partial\mathcal{P}^{out}}(\delta^TP\delta-\delta_0^TP\delta_0-2{\delta^*}^TP(\delta-\delta_0))
\nonumber \\& =\mathop
{\min}\limits_{\delta \in
\partial\mathcal{P}^{out}}(\delta^TP\delta-2{\delta^*}^TP(\delta-\delta_0)) -\delta_0^TP\delta_0
\end{align}

Hence, if $\mathop
{\min}\limits_{\delta \in
\partial\mathcal{P}^{out},\delta^*\in \Delta(\gamma)}(\delta^TP\delta-2{\delta^*}^TP(\delta-\delta_0)) >\delta_0^TP\delta_0,$
then $V_{\min}>V(x_0)$ for all $\delta^*\in \Delta(\gamma).$
Applying Theorem \ref{thr.StabilityAssessment}, we have Theorem
\ref{thr.RobustStability1} directly.  $\square$

\subsection{Proof of Theorem \ref{thr.RobustClearingTimeCertificate}}
\label{appen.RobustClearingTimeCertificate}

Similar to the proof of  Lemma \ref{thr.QuadraticLyapunov}, we have the derivative of $V(x)$ along the fault-on trajectory \eqref{eq.Faulton} as follows:
\begin{align}
\dot{V}(x_F) &= \dot{x_F}^TPx_F + x_F^TP\dot{x_F}  = x_F^T(A^TP+PA)x_F\nonumber\\& -2 x_F^TPBF + 2x_F^T PBD_{\{u,v\}}\sin\delta_{F_{uv}} \nonumber \\
&=W(x_F) -S^TS  + 2x_F^T PBD_{uv}\sin\delta_{F_{uv}} \nonumber \\
& +x_F^T\big[A^TP+PA- C^TK_1^TK_2C + R^TR \big]x_F
\end{align}
On the other hand
\begin{align}
2x_F^T PBD_{\{u,v\}}\sin\delta_{F_{uv}} &\le \mu x_F^T PBD_{\{u,v\}}D_{\{u,v\}}^TB^TP x_F \nonumber\\& +\dfrac{1}{\mu}\sin^2\delta_{F_{uv}}.
\end{align}
Therefore,

\begin{align}
\dot{V}(x_F)\le W(x_F) -S^TS +x_F^T\tilde{Q}x_F +\frac{1}{\mu}\sin^2\delta_{F_{uv}}
\end{align}
where $\tilde{Q}=A^TP+PA- C^TK_1^TK_2C + R^TR +\mu PBD_{\{u,v\}}D_{\{u,v\}}^TB^TP$. Note that $W(x_F)\le 0, \forall x_F\in \mathcal{P},$ and
\begin{align}
\tilde{Q}&=\bar{A}^TP+P\bar{A} +  \dfrac{(1-g)^2}{4}C^TC \nonumber \\&   + PBB^TP +\mu PBD_{\{u,v\}}D_{\{u,v\}}^TB^TP \le 0.
\end{align}
Therefore,
\begin{align}
\label{eq.Vfault}
\dot{V}(x_F) \le \frac{1}{\mu}\sin^2\delta_{F_{uv}}\le \frac{1}{\mu},
\end{align}
whener $x_F$ in the polytope $\mathcal{P}.$

We will prove that the
fault-cleared state $x_F(\tau_{clearing})$ is still in the set
$\mathcal{R}.$ It is easy to see that the
flow-in boundary $\partial\mathcal{P}^{in}$
prevents the fault-on dynamics \eqref{eq.Faulton} from escaping $\mathcal{R}.$

Assume that $x_F(\tau_{clearing})$ is not in the set
$\mathcal{R}.$ Then the fault-on trajectory can only escape
$\mathcal{R}$ through the segments which belong to sublevel set of
the Lyapunov function $V(x).$ Denote $\tau$ be the first time at
which the fault-on trajectory meets one of the boundary segments
which belong to sublevel set of the Lyapunov function $V(x).$
Hence $x_F(t) \in \mathcal{R}$ for all $0 \le t \le \tau.$ From
\eqref{eq.Vfault} and the fact that
$\mathcal{R}\subset \mathcal{P},$ we have

\begin{align}
V(x_F(\tau))-V(x_F(0)) = \int_0^{\tau} \dot{V}(x_F(t))dt \le
\frac{\tau}{\mu}
\end{align}
Note that $x_F(0)$ is the pre-fault equilibrium point, and thus equals to post-fault equilibrium point.
Hence, $V(x_F(0))=0$
and $\tau \ge \mu V(x_F(\tau)).$ By definition, we have
$V(x_F(\tau))=V_{\min}.$ Therefore, $\tau \ge \mu V_{\min},$
and thus, $\tau_{clearing}\ge \mu V_{\min},$ which is a
contradiction.  $\square$

\section{Acknowledgements}
This work was partially supported by NSF (Award No. 1508666),
MIT/Skoltech, Masdar initiatives and The Ministry of Education and
Science of Russian Federation (grant No. 14.615.21.0001, grant
code: RFMEFI61514X0001). The authors thank Dr. Munther Dahleh for his inspiring discussions motivating us to pursue problems in this paper.
We thank the anonymous reviewers for their careful reading of our manuscript and their many valuable comments and constructive suggestions.

\bibliographystyle{IEEEtran}
\bibliography{lff}

\newcommand{\noopsort}[1]{} \newcommand{\printfirst}[2]{#1}
  \newcommand{\singleletter}[1]{#1} \newcommand{\switchargs}[2]{#2#1}
\begin{thebibliography}{10}
\providecommand{\url}[1]{#1}
\csname url@samestyle\endcsname
\providecommand{\newblock}{\relax}
\providecommand{\bibinfo}[2]{#2}
\providecommand{\BIBentrySTDinterwordspacing}{\spaceskip=0pt\relax}
\providecommand{\BIBentryALTinterwordstretchfactor}{4}
\providecommand{\BIBentryALTinterwordspacing}{\spaceskip=\fontdimen2\font plus
\BIBentryALTinterwordstretchfactor\fontdimen3\font minus
  \fontdimen4\font\relax}
\providecommand{\BIBforeignlanguage}[2]{{%
\expandafter\ifx\csname l@#1\endcsname\relax
\typeout{** WARNING: IEEEtran.bst: No hyphenation pattern has been}%
\typeout{** loaded for the language `#1'. Using the pattern for}%
\typeout{** the default language instead.}%
\else
\language=\csname l@#1\endcsname
\fi
#2}}
\providecommand{\BIBdecl}{\relax}
\BIBdecl

\bibitem{1705631}
\textcolor{black}{Blaabjerg, F. and Teodorescu, R. and Liserre, M. and Timbus,
  A.V.}, ``\textcolor{black}{Overview of Control and Grid Synchronization for
  Distributed Power Generation Systems},'' \emph{\textcolor{black}{Industrial
  Electronics, IEEE Transactions on}}, vol. \textcolor{black}{53}, no.
  \textcolor{black}{5}, pp. \textcolor{black}{1398--1409},
  \textcolor{black}{Oct} \textcolor{black}{2006}.

\bibitem{turitsyn2011options}
\textcolor{black}{Turitsyn, Konstantin and Sulc, Petr and Backhaus, Scott and
  Chertkov, Michael}, ``\textcolor{black}{Options for control of reactive power
  by distributed photovoltaic generators},''
  \emph{\textcolor{black}{Proceedings of the IEEE}}, vol.
  \textcolor{black}{99}, no. \textcolor{black}{6}, pp.
  \textcolor{black}{1063--1073}, \textcolor{black}{2011}.

\bibitem{5454394}
\textcolor{black}{Rahimi, F. and Ipakchi, A.}, ``\textcolor{black}{Demand
  Response as a Market Resource Under the Smart Grid Paradigm},''
  \emph{\textcolor{black}{Smart Grid, IEEE Transactions on}}, vol.
  \textcolor{black}{1}, no. \textcolor{black}{1}, pp.
  \textcolor{black}{82--88}, \textcolor{black}{June} \textcolor{black}{2010}.

\bibitem{SPS}
P.~Anderson and B.~LeReverend, ``Industry experience with special protection
  schemes,'' \emph{Power Systems, IEEE Transactions on}, vol.~11, no.~3, pp.
  1166--1179, Aug 1996.

\bibitem{Huang:2012il}
Z.~Huang, S.~Jin, and R.~Diao, ``{Predictive Dynamic Simulation for Large-Scale
  Power Systems through High-Performance Computing},'' \emph{High Performance
  Computing, Networking, Storage and Analysis (SCC), 2012 SC Companion}, pp.
  347--354, 2012.

\bibitem{Nagel:2013kf}
I.~Nagel, L.~Fabre, M.~Pastre, F.~Krummenacher, R.~Cherkaoui, and M.~Kayal,
  ``{High-Speed Power System Transient Stability Simulation Using Highly
  Dedicated Hardware},'' \emph{Power Systems, IEEE Transactions on}, vol.~28,
  no.~4, pp. 4218--4227, 2013.

\bibitem{Pai:1981dv}
M.~A. Pai, K.~R. Padiyar, and C.~RadhaKrishna, ``{Transient Stability Analysis
  of Multi-Machine AC/DC Power Systems via Energy-Function Method},''
  \emph{Power Engineering Review, IEEE}, no.~12, pp. 49--50, 1981.

\bibitem{Chiang:2011eo}
H.-D. Chiang, \emph{{Direct Methods for Stability Analysis of Electric Power
  Systems}}, ser. Theoretical Foundation, BCU Methodologies, and
  Applications.\hskip 1em plus 0.5em minus 0.4em\relax Hoboken, NJ, USA: John
  Wiley {\&} Sons, Mar. 2011.

\bibitem{Chiang:1994ir}
H.-D. Chiang, F.~F. Wu, and P.~P. Varaiya, ``{A BCU method for direct analysis
  of power system transient stability},'' \emph{Power Systems, IEEE
  Transactions on}, vol.~9, no.~3, pp. 1194--1208, Aug. 1994.

\bibitem{Tong:2010}
J.~Tong, H.-D. Chiang, and Y.~Tada, ``{On-line power system stability screening
  of practical power system models using TEPCO-BCU},'' in \emph{ISCAS}, 2010,
  pp. 537--540.

\bibitem{Hill:1989:LFL:72068.72082}
D.~J. Hill and C.~N. Chong, ``Lyapunov functions of {L}ur'e-{P}ostnikov form
  for structure preserving models of power systems,'' \emph{Automatica},
  vol.~25, no.~3, pp. 453--460, May 1989.

\bibitem{Hiskens:1997Lya}
R.~Davy and I.~A. Hiskens, ``{Lyapunov functions for multi-machine power
  systems with dynamic loads},'' \emph{Circuits and Systems I: Fundamental
  Theory and Applications, IEEE Transactions on}, vol.~44, no.~9, pp. 796--812,
  1997.

\bibitem{hiskens2006sensitivity}
I.~A. Hiskens and J.~Alseddiqui, ``Sensitivity, approximation, and uncertainty
  in power system dynamic simulation,'' \emph{Power Systems, IEEE Transactions
  on}, vol.~21, no.~4, pp. 1808--1820, 2006.

\bibitem{dong2012numerical}
Z.~Y. Dong, J.~H. Zhao, and D.~J. Hill, ``Numerical simulation for stochastic
  transient stability assessment,'' \emph{Power Systems, IEEE Transactions on},
  vol.~27, no.~4, pp. 1741--1749, 2012.

\bibitem{dhople2013analysis}
S.~V. Dhople, Y.~C. Chen, L.~DeVille, and A.~D. Dom{\'\i}nguez-Garc{\'\i}a,
  ``Analysis of power system dynamics subject to stochastic power injections,''
  \emph{Circuits and Systems I: Regular Papers, IEEE Transactions on}, vol.~60,
  no.~12, pp. 3341--3353, 2013.

\bibitem{VuTuritsyn:2014}
T.~L. Vu and K.~Turitsyn, ``Lyapunov functions family approach to transient
  stability assessment,'' \emph{IEEE Transactions on Power Systems}, vol.~31,
  no.~2, pp. 1269--1277, March 2016.

\bibitem{OPT-006}
\BIBentryALTinterwordspacing
L.~Vandenberghe and M.~S. Andersen, ``Chordal graphs and semidefinite
  optimization,'' \emph{Foundations and Trends in Optimization}, vol.~1, no.~4,
  pp. 241--433, 2014. [Online]. Available:
  \url{http://dx.doi.org/10.1561/2400000006}
\BIBentrySTDinterwordspacing

\bibitem{VuTuritsyn:2014acc}
T.~L. Vu and K.~Turitsyn, ``Synchronization stability of lossy and uncertain
  power grids,'' in \emph{2015 American Control Conference (ACC)}, July 2015,
  pp. 5056--5061.

\bibitem{VuTuritsyn:2014pes}
------, ``Geometry-based estimation of stability region for a class of
  structure preserving power grids,'' in \emph{2015 IEEE Power Energy Society
  General Meeting}, July 2015, pp. 1--5.

\bibitem{zhao2014design}
C.~Zhao, U.~Topcu, N.~Li, and S.~Low, ``Design and stability of load-side
  primary frequency control in power systems,'' \emph{Automatic Control, IEEE
  Transactions on}, vol.~59, no.~5, pp. 1177--1189, 2014.

\bibitem{mallada2014optimal}
E.~Mallada, C.~Zhao, and S.~Low, ``Optimal load-side control for frequency
  regulation in smart grids,'' in \emph{Communication, Control, and Computing
  (Allerton), 2014 52nd Annual Allerton Conference on}, Sept 2014, pp.
  731--738.

\bibitem{Backhaus:2014}
S.~Backhaus, R.~Bent, D.~Bienstock, M.~Chertkov, and D.~Krishnamurthy,
  ``{Efficient synchronization stability metrics for fault clearing},''
  \emph{Available: arXiv:1409.4451}.

\bibitem{Anghel:2013}
M.~Anghel, J.~Anderson, and A.~Papachristodoulou, ``{Stability analysis of
  power systems using network decomposition and local gain analysis},'' in
  \emph{2013 IREP Symposium-Bulk Power System Dynamics and Control}.\hskip 1em
  plus 0.5em minus 0.4em\relax IEEE, 2013, pp. 978--984.

\bibitem{ortega2005transient}
R.~Ortega, M.~Galaz, A.~Astolfi, Y.~Sun, and T.~Shen, ``Transient stabilization
  of multimachine power systems with nontrivial transfer conductances,''
  \emph{Automatic Control, IEEE Transactions on}, vol.~50, no.~1, pp. 60--75,
  2005.

\bibitem{galaz2003energy}
M.~Galaz, R.~Ortega, A.~S. Bazanella, and A.~M. Stankovic, ``An energy-shaping
  approach to the design of excitation control of synchronous generators,''
  \emph{Automatica}, vol.~39, no.~1, pp. 111--119, 2003.

\bibitem{shen2000adaptive}
T.~Shen, R.~Ortega, Q.~Lu, S.~Mei, and K.~Tamura, ``Adaptive l 2 disturbance
  attenuation of hamiltonian systems with parametric perturbation and
  application to power systems,'' in \emph{Decision and Control, 2000.
  Proceedings of the 39th IEEE Conference on}, vol.~5.\hskip 1em plus 0.5em
  minus 0.4em\relax IEEE, 2000, pp. 4939--4944.

\bibitem{bergen1981structure}
A.~R. Bergen and D.~J. Hill, ``A structure preserving model for power system
  stability analysis,'' \emph{Power Apparatus and Systems, IEEE Transactions
  on}, no.~1, pp. 25--35, 1981.

\bibitem{Dorfler:2013}
F.~Dorfler, M.~Chertkov, and F.~Bullo, ``{Synchronization in complex oscillator
  networks and smart grids},'' \emph{Proceedings of the National Academy of
  Sciences}, vol. 110, no.~6, pp. 2005--2010, 2013.

\bibitem{Popov:1962}
V.~M. Popov, ``{Absolute stability of nonlinear systems of automatic
  control},'' \emph{Automation Remote Control}, vol.~22, pp. 857--875, 1962,
  Russian original in Aug. 1961.

\bibitem{Yakubovich:1967}
V.~A. Yakubovich, ``{Frequency conditions for the absolute stability of control
  systems with several nonlinear or linear nonstationary units},''
  \emph{Avtomat. i Telemekhan.}, vol.~6, pp. 5--30, 1967.

\bibitem{Megretski:1997}
A.~Megretski and A.~Rantzer, ``{System analysis via integral quadratic
  constraints},'' \emph{Automatic Control, IEEE Transactions on}, vol.~42,
  no.~6, pp. 819--830, 1997.

\bibitem{118bus}
{https://www.ee.washington.edu/research/pstca/pf118/pg\_tca118bus.htm}.

\bibitem{Zhong_Synchronverters}
\textcolor{black}{Qing-Chang Zhong and Weiss, G.},
  ``\textcolor{black}{Synchronverters: Inverters That Mimic Synchronous
  Generators},'' \emph{\textcolor{black}{Industrial Electronics, IEEE
  Transactions on}}, vol. \textcolor{black}{58}, no. \textcolor{black}{4}, pp.
  \textcolor{black}{1259--1267}, \textcolor{black}{April}
  \textcolor{black}{2011}.

\bibitem{Kundur}
P.~Kundur, \emph{{Power System Stability and Control}}, New York, 1994.

\bibitem{boyd1994linear}
S.~Boyd, L.~El~Ghaoui, E.~Feron, and V.~Balakrishnan, \emph{Linear matrix
  inequalities in system and control theory}.\hskip 1em plus 0.5em minus
  0.4em\relax SIAM, 1994, vol.~15.

\bibitem{januszquiros2014constrained}
J.~Quir{\'o}s-Tort{\'o}s, R.~S{\'a}nchez-Garc{\'\i}a, J.~Brodzki, J.~Bialek,
  and V.~Terzija, ``Constrained spectral clustering-based methodology for
  intentional controlled islanding of large-scale power systems,'' \emph{IET
  Generation, Transmission \& Distribution}, vol.~9, no.~1, pp. 31--42, 2014.

\bibitem{gan2000stability}
D.~Gan, R.~J. Thomas, and R.~D. Zimmerman, ``Stability-constrained optimal
  power flow,'' \emph{Power Systems, IEEE Transactions on}, vol.~15, no.~2, pp.
  535--540, 2000.

\bibitem{yuan2003solution}
Y.~Yuan, J.~Kubokawa, and H.~Sasaki, ``A solution of optimal power flow with
  multicontingency transient stability constraints,'' \emph{Power Systems, IEEE
  Transactions on}, vol.~18, no.~3, pp. 1094--1102, 2003.

\end{thebibliography}

\begin{IEEEbiography}[{\includegraphics[width=1in,height=1.25in,clip,keepaspectratio]{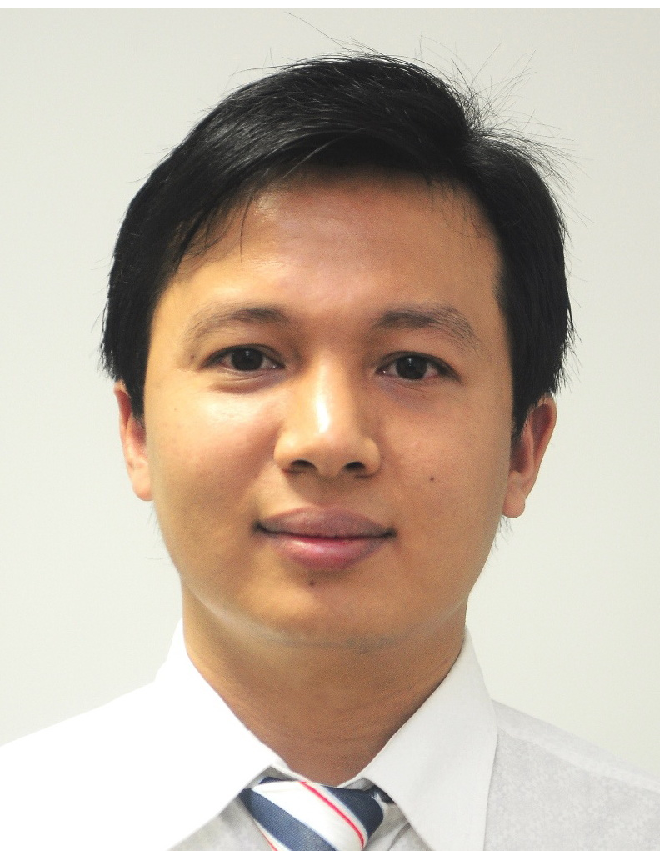}}]{Thanh Long Vu} received the B.Eng. degree in automatic control from Hanoi University of Technology in 2007 and the Ph.D. degree in electrical engineering from National University of Singapore in 2013.  Currently, he is a Research Scientist at the Mechanical Engineering Department of Massachusetts Institute of Technology (MIT). Before joining MIT, he was a Research Fellow at Nanyang Technological University, Singapore. His main research interests lie at the intersections of electrical power systems, systems theory, and optimization. He is currently interested in exploring robust and computationally tractable approaches for risk assessment, control, management, and design of large-scale complex systems with emphasis on next-generation power grids.
\end{IEEEbiography}
\begin{IEEEbiography}[{\includegraphics[width=1in,height=1.25in,clip,keepaspectratio]{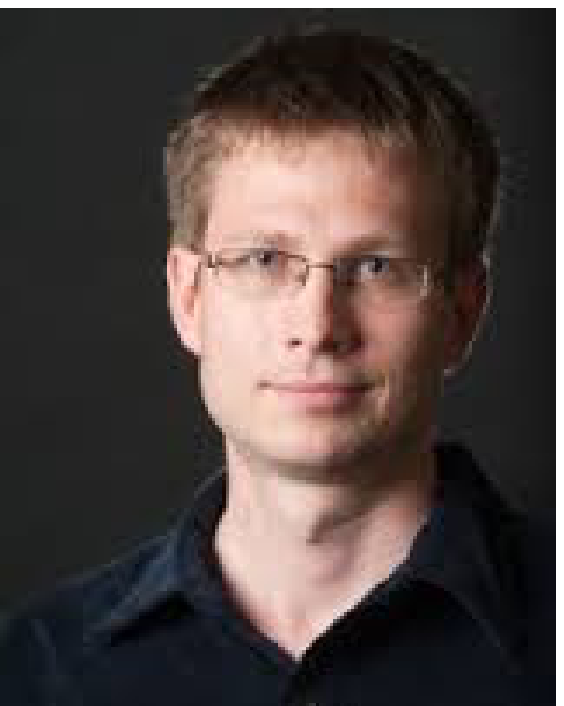}}]{Konstantin Turitsyn} (M`09) received the M.Sc. degree in physics from Moscow Institute of Physics and Technology and the Ph.D. degree in physics from Landau Institute for Theoretical Physics, Moscow, in 2007.  Currently, he is an Assistant Professor at the Mechanical Engineering Department of Massachusetts Institute of Technology (MIT), Cambridge. Before joining MIT, he held the position of Oppenheimer fellow at Los Alamos National Laboratory, and Kadanoff–Rice Postdoctoral Scholar at University of Chicago. His research interests encompass a broad range of problems involving nonlinear and stochastic dynamics of complex systems. Specific interests in energy related fields include stability and security assessment, integration of distributed and renewable generation. He is the recipient of the 2016 NSF CAREER award.
\end{IEEEbiography}
\end{document}